\documentclass[prl,aps,twocolumn,showpacs,superscriptaddress]{revtex4-1}

\usepackage{graphicx}
\usepackage{dcolumn}
\usepackage{bm}
\usepackage{dsfont}
\usepackage{amssymb,amsmath}
\usepackage{sidecap}
\usepackage{wrapfig}

\usepackage{graphicx}
\usepackage{leftidx}

\newcommand{\ket}[1]{\left| #1 \right>} 
\newcommand{\bra}[1]{\left< #1 \right|} 
\newcommand{\vev}[1]{\langle #1 \rangle}
\newcommand{\up}{\uparrow}
\newcommand{\dw}{\downarrow}
\newcommand{\id}{\mathbb{I}}

\begin{document}

\title{Non-classical correlations in a class of spin chains with long-range interactions and exactly solvable ground states}

\author{Emanuele Levi}
\affiliation
{School of Physics and Astronomy, University of Nottingham, Nottingham, NG7 2RD, United Kingdom}
\author{Igor Lesanovsky}
\affiliation
{School of Physics and Astronomy, University of Nottingham, Nottingham, NG7 2RD, United Kingdom}

\begin{abstract}
We introduce a class of spin models with long-range interactions---in the sense that they extend significantly beyond nearest neighbors---whose ground states can be constructed analytically and have a simple matrix
product state representation. This enables the detailed study of ground state properties, such as correlation functions and entanglement, in the thermodynamic limit. The spin models presented here are closely related to lattice gases of strongly interacting polar molecules or Rydberg atoms which feature an excluded volume or blockade interaction. While entanglement is only present between spins that are separated by no more than a blockade length, we show that non-classical correlations can extend much further and analyze them through quantum discord.
We furthermore identify a set of seemingly critical points where the ground state approaches a crystalline state with a filling fraction that is given by the inverse of the blockade length. We analyze the scaling properties in the vicinity of this parameter region and show that the correlation length possesses a non-trivial dependence on the blockade length.
\end{abstract}

\maketitle
\textit{Introduction.}-- Finding exact ground states of quantum many-body Hamiltonians with interactions that extend far beyond nearest neighbors is typically a very challenging tasks in condensed matter physics. This is because exactly solvable cases, e.g. the Haldane-Shastry model \cite{haldane}, are extremely rare, and the numerical treatment of long-range interactions is computationally demanding even with modern numerical tools such as the Density Matrix Renormalization Group \cite{dmrg1,dmrg2}.\\
From the experimental side ensembles of cold atoms, ions and molecules offer a very promising route towards the controlled study of long-range
interactions in quantum many-body systems. Very recent experimental approaches employ atoms in highly excited states --- so-called Rydberg atoms --- where they exhibit strong dipolar interactions. A characteristic feature of these systems is the presence of the dipole blockade which prevents the excitation of an atom in the vicinity of an already excited one \cite{blockade1,blockade2}. In typical experimental setups one can achieve situations in which a single excitation blocks tens or hundreds of atoms and can in this sense be regarded as long ranged. In the extreme case the blockade can extend over an entire cloud which leads to the formation of so-called ``super atoms'' --- an entangled state of a single delocalized excitation \cite{kuzmich}. Of particular interest is the more involved case in which the system size is larger than a blockade region. Clearly, the blockade leads to a strong anti-correlation of excitations at small distances. However, the nature of the correlations at longer distances is at present not fully understood. A number of recent investigations suggest that the emerging states behave essentially classically in the sense that their properties can be understood by drawing an analogy to arrangements of classical hard objects \cite{Pohl2010,schachenmayer,petrosyan,cenap1,Garttner12,Schauss12,Petrosyan13}. To what extent there exist quantum correlations that go beyond the aforementioned ``super atom'' states is so far unclear.\\
\begin{figure}[h]
\includegraphics[width=\columnwidth]{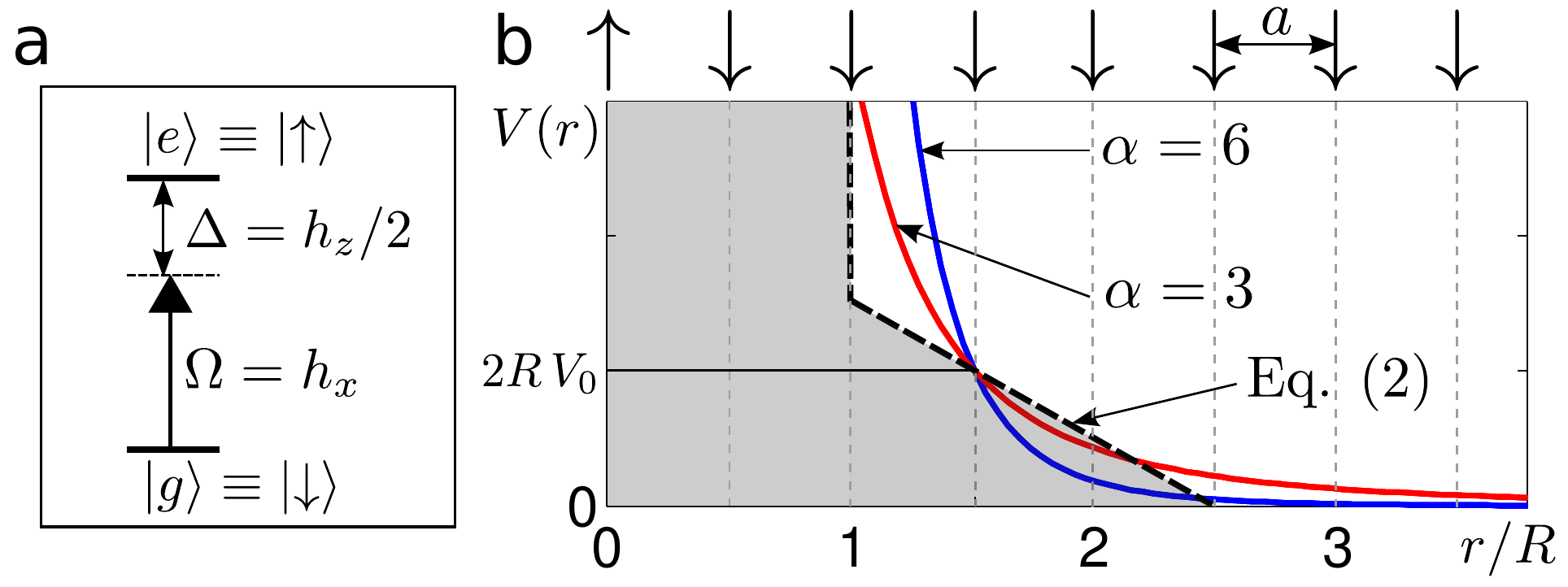}
\caption{(a) The spin system described by Hamiltonian (\ref{eq:ham}) is closely related to ensembles of interacting Rydberg atoms or polar molecules. They can be modelled by two-level systems whose excited state $\left|e\right>$ and ground state $\left|g\right>$ are coupled by a laser or microwave field of strength $\Omega$ and detuning $\Delta$. (b) Spins in the up-state interact with the interaction potential $V(r)$ [Eq. \ref{eq:pot}] which can be regarded as an approximation of power-law potential of the form $V_\alpha(r)=C_\alpha/(a\,r )^\alpha$ where $a$ is the lattice spacing. For further explanation see text.}
\label{fig:potential}
\end{figure}
One motivation of this paper is conduct a largely analytical study to shed some light on these questions. A second one is to introduce a class of long-range interacting one-dimensional spin models whose ground state in some regime can be solved exactly. The models, which are all of Ising type, manifestly display the blockade effect due to an excluded volume interaction that encompasses $R$ spins. Moreover, they possess a potential tail which extends further than $R$ and therefore mimic to a good extent the typical features present in strongly interacting Rydberg gases. In the exactly solvable regime their ground state has the form of a matrix product state, which permits the convenient calculation of the correlation properties in the thermodynamic limit. This unique property allows us to perform a scaling study of the correlation length which is shown to exhibit a non-trivial power-law dependence on $R$. We find that the expectation values of classical observables can indeed be understood from analogous classical arrangements of hard objects, and that entanglement between two spins is only present when they are separated by at most one blockade length $R$. When separated further, despite the absence of entanglement, non-classical correlations remain in form of quantum discord \cite{discord1,discord2}, which is regarded as key resource for conducting quantum operations in the presence of noise such as quantum illumination \cite{qillum1,qillum2,qillum3}, and metrology with noisy probes \cite{noisymetrology,discord3}.
The fact that in the systems studied here quantum correlations extend over distances larger than $R$ also hints towards the possibility to implement non-classical operations between distant particles mediated by Rydberg interactions in experimental ensembles which are not fully blockaded.

\textit{Hamiltonian.} --
The class of Hamiltonians we are considering is that of one dimensional lattice spin-$\frac{1}{2}$ models with transverse and longitudinal magnetic
fields and an Ising-type interaction potential:
\begin{equation}
\label{eq:ham}
H_0=\sum_k^N \left(h_x \sigma^x_k+h_z\sigma^z_k+\sum_{m>k}V_{km}n_k n_m\right).
\end{equation}
Here $\sigma^x$ and $\sigma^z$ are Pauli matrices and $n=(\id+\sigma^z)/2$.
The interaction energy $V_{km}$ between spins positioned at sites $k$ and $m$ is given by the potential $V(|k-m|)$ with
\begin{equation}
\label{eq:pot}
V(r)=\left\{
  \begin{array}{l l}
    \infty & \quad \text{if $|r|\leq R$, }\\
    V_0\times [2R-(|r|-1)] & \quad \text{if $R < |r| \leq 2R $, }\\
    0 & \quad \text{if $|r| > 2R $.}
  \end{array} \right.
\end{equation}
It features a hard core interaction between up-spins up to a distance $R$. Beyond that the potential decays linearly until it reaches the distance $2R$
from where onwards it is zero. With this potential it is energetically forbidden to dynamically access configurations in which the separation between any two spins is smaller than $R$ sites.

Such potential can be linked to current studies of strongly interacting lattice gases of cold Rydberg atoms or polar molecules \cite{schachenmayer}. These systems can be described in terms of ensembles of interacting two-level systems with the ground state $\ket{g}\equiv\ket{\dw}$, and and excited state $\ket{e}\equiv\ket{\up}$. The transition between the two levels is driven by a coherent laser or microwave field with detuning $\Delta$ and Rabi frequency $\Omega$, as shown in Fig.~\ref{fig:potential}a. The single spin terms in Eq.~(\ref{eq:ham}) correspond the Hamiltonian of non-interacting driven two-level systems when setting $h_x=\Omega$ and $h_z=\Delta/2$. The interaction between excited atomic or molecular states typically decays as a power-law $V_\alpha(r)=C_\alpha/(a\,r)^{\alpha}$ with power $\alpha$ being $3$ or $6$ and $a$ being the lattice spacing. Due to this interaction certain spin configurations become dynamically inaccessible. In particular the simultaneous excitation of two particles is strongly suppressed if their interaction energy is larger than the value of the Rabi frequency. This defines a blockade length $R_\mathrm{b}\sim (C_\alpha/\Omega)^{1/\alpha}$ \cite{Sun08} which can be identified with the parameter $R$ in $V(r)$ [Eq. (\ref{eq:pot})]. Due to the power-law decay the interaction potentials extend beyond $R_\mathrm{b}$. These tails can be thought of being mimicked by the linearly decaying part of $V(r)$. To approximately connect $V(r)$ to the power-law potentials $V_\alpha(r)$ by setting $V(R+1)=V_\alpha(R+1)$. With this we find $V_0=C_\alpha/(R [a\,(R+1)]^\alpha)$. A comparison of these potentials is shown in Fig. \ref{fig:potential}b.

\textit{Exactly solvable parameter manifold.} -- As $V(r)$ forbids the simultaneous excitation of spins at distances closer or equal than $R$ the physically relevant subspace of the Hilbert space is spanned by all states $\ket{\psi_\nu}$ which obey $n_kn_{k+1}\ket{\psi_\nu}=n_kn_{k+2}\ket{\psi_\nu}=...=n_kn_{k+R}\ket{\psi_\nu}=0$. Within this physical sector it can be shown that the Hamiltonian (\ref{eq:ham}) acquires a frustration free or Rokhsar-Kivelson form \cite{Rokhsar}, provided that the system parameters obey
\begin{eqnarray}
  h_z=\frac{1}{2}\left[\frac{h_x^2}{V_0}-V_0(2R+1)\right].\label{eq:RK_manifold}
\end{eqnarray}
Specifically, on this exactly solvable manifold Eq. (\ref{eq:ham}) can be brought into the form $H=E_0+\sum_k^N \mathcal{H}_k$ with
\begin{eqnarray}
  \mathcal{H}_k=h_x  \sum^N_k \mathcal{P}^\mathrm{L}_k\left[\sigma^x_k+\frac{h_x}{V_0}n_k+\frac{V_0}{h_x} P_k\right]\mathcal{P}^\mathrm{R}_k.
  \label{eq:RK_Hamiltonian}
\end{eqnarray}
Here we have abbreviated the string operators $\mathcal{P}^\mathrm{L}_k=P_{k-1}P_{k-2}...P_{k-R}$,
and $\mathcal{P}^\mathrm{R}_k=P_{k+1}P_{k+2}...P_{k+R}$ which are products of the projector $P_k=1-n_k$ that projects on the spin-down state of the $k$-th spin. This Hamiltonian is a generalization of the ones presented in Refs. \cite{Fendley04,Lesanovsky11,Lesanovsky12,Katsura13}. It is composed by local positive-semidefinite Hamiltonians $\mathcal{H}_k$, which in general do not commute. They all annihilate the ground state $\ket{z}$ (see further below for discussion), i.e. $\mathcal{H}_k\ket{z}=0$, and hence the ground state energy on the parameter manifold (\ref{eq:RK_manifold}) is given by $E_0=-N(h_z+V_0)$.

\textit{Ground state wave function and correlations}--
 \begin{figure}
\includegraphics[width=\columnwidth]{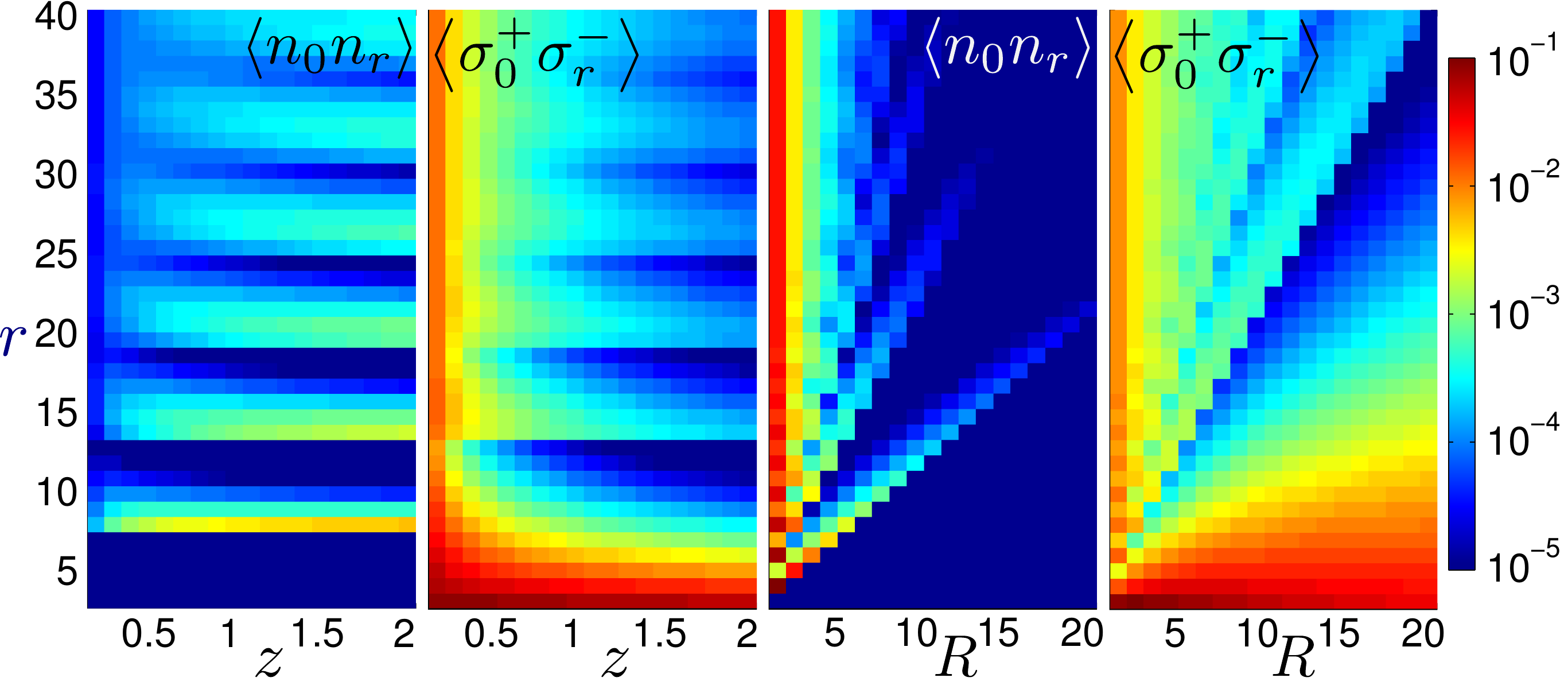}\caption{From left to right: Density-density correlation and spatial coherence for $R=5$. Density-density correlation and spatial coherence for $z=5$.} \label{fig:corr}
 \end{figure}
The ground state wave function can be explicitly written as
\begin{align}
\ket{z}=\frac{\prod_{k=1}^N e^{-z\mathcal{P}^\mathrm{L}_k\sigma^x_k \mathcal{P}^\mathrm{R}_k}}{\sqrt{Z(z,N)}}\ket{0},
 \label{eq:gs}
\end{align}
where $\ket{0}=\ket{\dw\dw...\dw}$ is the spin vacuum. The state $\ket{z}$ is a superposition of all classical spin configurations in which up-spins are at least separated by a distance $R$. The relative weight of each configuration is given by $z^{2m}$, with the parameter $z=V_0/h_x$ (which we take to be positive in the following) and $m$ being the number of up-spins contained in the configuration. The state space is equivalent to that of hard $R+1$-mers on a lattice and hence the normalization constant $Z(z,N)$ is given by the classical grand-canonical partition function of hard
$R+1$-mers with fugacity $z^2$.

For given $R$ the ground state in Eq.~(\ref{eq:gs}) assumes an exact matrix product state (MPS) form \cite{mps1,mps2}, that is $\ket{z}=\sum_{\{i_1,...,i_N\}=\downarrow,\uparrow}\psi_z(i_1,...,i_N)\ket{i_1,...,i_N}$ with $\psi_z(i_1,...,i_N)=\mathrm{Tr}\left[ \, X_{i_1}X_{i_2}...X_{i_L}\right]$. Here $X_\uparrow = \delta_{1,R+1}$ and $X_\downarrow = \delta_{R,R}-z\delta_{R+1,R}+\sum_{m=1}^{R-1}\delta_{m+1,m}$ are $(R+1)\times(R+1)$-dimensional matrices. With the MPS representation it is a relatively simple task to characterize the properties of the ground state, e.g. its correlation functions and entanglement properties in the thermodynamic limit: To this end we define the transfer operator $ E^O = \sum_{i,i'=\uparrow,\downarrow} X_i \otimes X_{i'} \bra{i'} O \ket{i}$ such that the correlation functions
can be brought into the form $\vev{O_0 O'_r}=\sum_{\alpha=1}^{(R+1)^2} c_\alpha e^{-\frac{r}{\xi_\alpha}}\left( \cos \phi_\alpha+i \sin \phi_\alpha\right)$, where $c_\alpha=\bra{l_1}E^O\ket{r_\alpha} \bra{l_\alpha}E^{O'}\ket{r_1}$. The vectors $\bra{l_\alpha}$ and $\ket{r_\alpha}$ form the left
and right eigenbasis of the transfer operator $E^{\id}$, while $\xi^{-1}_\alpha=\log\left|\lambda_1/\lambda_\alpha\right|$, and
$\phi_\alpha=\arg\left( \lambda_1/\lambda_\alpha\right)$, where $|\lambda_\alpha|\geq |\lambda_{\alpha+1}|$ are the eigenvalues of
$E^{\id}$.\\
In Fig.~\ref{fig:corr} we display the density-density correlation function $\vev{n_0 n_r}$ and the spatial coherence $\vev{\sigma^{+}_0\sigma^{-}_r}$. The density-density correlation function exhibits decaying oscillations at a length scale that is approximately given by $R$. With increasing $R$, and keeping $z$ fixed, the amplitude of the oscillations decreases. Keeping $R$ constant and varying $z$, we observe that the oscillation become increasingly pronounced with growing $z$. In the limit of $z\rightarrow\infty$ configurations that contain the highest possible number of excitations (compatible with the blockade) carry almost all the weight, and the ground state approaches a superposition of $R+1$ ``crystalline'' states $\ket{c}_{m}$ each of which contains a regularly ordered arrangement of up-spins with nearest neighbor distance $R+1$ and the first up-spin being located at site $m$: $\ket{z\rightarrow \infty}_{R}=[\ket{c}_{1}+...+\ket{c}_{R+1}]\sqrt{R+1}$.
The correlation length of $\ket{z\rightarrow \infty}_{R}$ is infinite and in fact the two largest eigenvalues of the transfer operator $E^{\id}$ have the same magnitude when $z$ approaches infinity, which we from now on refer to as critical limit. The spatial coherence $\vev{\sigma^{+}_0\sigma^{-}_r}$ shows some qualitative analogies with the density-density correlation (see Fig.~\ref{fig:corr}). It is strongly decaying with increasing spin separation, with an oscillatory pattern whose contrast is more and more suppressed as $R$ increases. Opposite to the behaviour of the density-density correlation function the spatial coherence is more strongly suppressed the larger $z$, and vanishes at the critical point. These numerical results confirm the nature of the state $\ket{z\rightarrow \infty}_{R}$, for which one can show that the spatial coherence is identically zero, for any pair of spins.

To conclude the discussion on the correlations we consider the situation in which we keep the blockade length constant while decreasing the lattice spacing $a$. In practice this can be achieved experimentally in Rydberg (molecular) gases by increasing the density of atoms (molecules). This scenario is in fact interesting because recent studies of driven Rydberg gases \cite{cenap1,Garttner12,petrosyan} suggest that spatial correlations can become enhanced by increasing the atomic density. To study whether this also applies here we define the dimensionful blockade length $\tilde{R}=aR$ and introduce a continuous set of coordinates $x=ka$, by the help of which we can express the ground state (\ref{eq:gs}) as
\begin{equation}
 \label{eq:contstate}
 \begin{split}
 \ket{\tilde{z}}=\left[\Xi(\tilde{z},\tilde{R},\tilde{L})\right]^{-1/2}\sum_{n=0}^{\tilde{L}/\tilde{R}} \frac{(-\tilde{z})^{n}}{n!}\int_{0}^{\tilde{L}} dx_1...dx_n\\
 \psi(x_1,...,x_n)\phi^\dagger(x_1)...\phi^\dagger(x_n)\ket{0},
\end{split}
 \end{equation}
with $\psi(x_1,...,x_n)=\theta(x_n-x_{n-1}-\tilde{R})...\theta(x_2-x_{1}-\tilde{R})$, $\psi^\dagger(x)=\sigma^+_k/\sqrt{a}$, and $\tilde{z}=z/\sqrt{a}$, $\theta(x)$ being the step function. The normalization is $\Xi(\tilde{z},\tilde{R},\tilde{L})=\sum_{n=0}^{\tilde{L}/\tilde{R}}\tilde{z}^{2n} \xi(n,\tilde{R},\tilde{L})$, where $\xi(n,\tilde{R},\tilde{L})$ is the microcanonical partition function of a Tonks gas \cite{book}, i.e. of $n$ hard rods of length $\tilde{R}$ arranged in a system of length $\tilde{L}=aL$. The correlation length of the above state is controlled by the parameter $\tilde{z}$ which diverges as $a$ tends to zero. Hence, for fixed $z$ the density-density correlations become longer ranged when $a$ is decreased, i.e. the density is increased.
In order to define the state (\ref{eq:contstate}) with a finite $\tilde{R}$ we need to consider a diverging blockade length $R$.
In this limit the bond dimension of $\ket{z}$ becomes infinite, such that (\ref{eq:contstate}) stands as an example of a continuous limit of an MPS which is not expressible as a continuous matrix product state \cite{cmps1,cmps2}.

 \begin{figure*}
\includegraphics[width=1.8 \columnwidth]{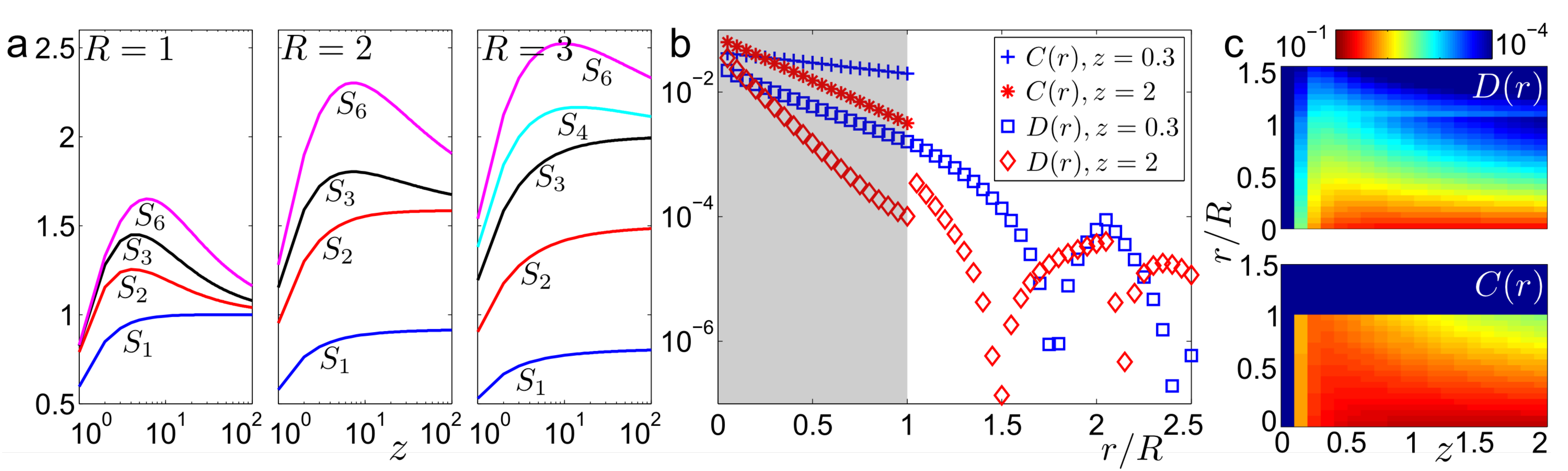}
\caption{(a) Entanglement entropy $S_r$ of spin blocks of various sizes $r$. If $r \leq R$ it is a monotonously increasing function while it possesses a maximum when the block size exceeds the blockade length $R$. (b) Concurrence $C(r)$ and quantum discord $D(r)$ for $R=20$ and $z=0.3, 2$. Beyond $r=R$ $C(r)$ is strictly zero. In contrast to that $D(r)$ is non-zero also in this region and exhibits an oscillatory behaviour.(c) Density plot of concurrence $C(r)$ and the quantum discord $D(r)$ for $R=20$. The sharp cut-off of entanglement (in contrast to quantum discord) at $r=R$ is clearly visible.}
        \label{fig:quantum}
 \end{figure*}
\textit{Entanglement and non-classicality}--
Due to the structure of the ground state (\ref{eq:gs}) the expectation value of classical observables, e.g. the density-density correlation function, is equivalent to that of classical hard $R+1$-mers with fugacity $z^2$ \cite{Lesanovsky11}. However, as we have shown before the ground state also exhibits quantum coherence.

We are therefore interested in the question as to whether it also features non-classical correlations, such as entanglement. To find an answer we start by considering two figures of merit of entanglement, namely the block entropy and the concurrence \cite{concurrence1,concurrence2}. The first captures the collective properties of entanglement of a block of a certain number of contiguous spins, while the second quantifies the entanglement shared by a pair of spins. The block entropy, defined as $S_{r}=-\text{Tr} \rho_{r}\log_2 \rho_{r}$, depends on the reduced density matrix
$\rho_{r}=\frac{1}{\lambda_1^r}\sum_{\{i_{j},i'_{j}\}}\text{Tr}\left[\tilde{B}\prod_{j=1}^{r}E_{i_{j},i'_j}\right]
\ket{i_{1},i_{2},...,i_{l}}\bra{i'_{1},i'_{2},...,i'_{l}}$,
where $\tilde{B}=\lim_{L\rightarrow\infty}\left( E^{\id}/\lambda_1\right)^{L}$, and $E_{i_{j},i'_j}=X_{i_j} \otimes X_{i'_j}$. The ground state (\ref{eq:gs}) factorizes at the point $z=0$ for any value of $R$, leading to zero entropy as a result, signaling an overall classical state. In the limit $z\rightarrow\infty$, on the other hand, the entropy approaches an asymptotic values, which can be extracted from the state $\ket{z\rightarrow \infty}_R$: $S_r(R)=\log_2(R+1)-\Theta(R-r+1)/(R+1)\log_2(R-r+1)$, where $\Theta(x)=x\theta(x)$.

In \cite{Lesanovsky11} the single atom entropy was considered in the case $R=1$, and it was found to decrease monotonically with increasing
$z$. This is not true for general block sizes and blockade lengths $R$, as shown in Fig.~\ref{fig:quantum}a: If $r\leq R$ the entropy is a monotonously increasing function in $z$. However, as soon as the block size exceeds the blockade length $S_r$ exhibits a maximum in $z$, whose precise location depends on $R$ and $r$. This qualitative change in the behavior is due to the fact that within the blockade length the state space is restricted to configurations with at most one up-spin whereas as soon as $r>R$ the number of accessible configurations grows fast therefore allowing for an entropy larger than the asymptotic value $S_r(R)$.

To quantify the entanglement between pairs of spins separated by a distance $r$ we study the concurrence which is defined as $C(r)=\text{max}\{2\lambda_1-\text{Tr}B,0\}$, where $\lambda_1$ is the largest eigenvalue of the matrix $B=\sqrt{\sqrt{\rho(r)}\tilde{\rho}(r)\sqrt{\rho(r)}}$.
Here $\rho(r)=\rho_{k,k+r}$ is the reduced density matrix of the two spins, and $\tilde{\rho}(r)$ is the this matrix expressed in the Bell basis \cite{bell}. The concurrence is plotted Fig.~\ref{fig:quantum}b (cut for $z=0.3$ and $z=2$ at fixed $R=20$) as well as in the bottom panel of Fig.~\ref{fig:quantum}c. Clearly there is no entanglement shared by two spins separated by a distance $r>R$, since here $C(r)$ drops sharply to zero. Hence the blockade length $R$ is equal to the range of entanglement. Linking back to the systems of interacting Rydberg atoms this shows that entanglement indeed only extends over the size of a ``super atom''.

Entanglement though does not represent all possible quantum correlations between two spins. They are instead captured by the quantum discord which we study in the following. We use the local quantum uncertainty \cite{discord3} as a measure of discord, defined as $D(r)=1-\Lambda_\text{max}$, where $\Lambda_\text{max}$ is the largest eigenvalue of the $3\times 3$ matrix of entries $W_{ij}=\text{Tr}\left[\sqrt{\rho(r)}(\sigma^i\otimes \id)\sqrt{\rho(r)}(\sigma^j\otimes \id)\right]$.  The discord is shown in Fig.~\ref{fig:quantum}b (cut for $z=0.3$ and $z=2$ at fixed $R=20$) and in the top panel of Fig.~\ref{fig:quantum}c. Surprisingly, quantum correlations in form of discord extend much further than entanglement. Furthermore, it is interesting to note that quantum discord shows an actual oscillatory behavior [Fig.~\ref{fig:quantum}b] as function of $r$.

\textit{Critical limit.} -- As discussed previously the limit $z\rightarrow\infty$ can be thought of as a critical limit where in fact all ``crystalline configurations'' $\ket{c}_m$ are valid ground states. Translational symmetry is broken as the states $\ket{c}_m$ are only invariant under translations by $R$ sites. The formation and melting of such crystalline states realized in a one-dimensional gas of interacting Rydberg atoms has been investigated in Ref. \cite{devil}. Here the authors identified a ``devils stair case'' in the phase diagram formed by ``crystals'' with different filling fraction. Linking to this study, we can now actually understand how within our model the correlation length $\xi$ diverges as $z$ approaches infinity, i.e. when the crystal is formed. Interestingly, this does strongly depend on the value of $R$. For all blockade lengths $\xi$ diverges with a characteristic power law, $\xi_R\sim z^{\nu_R}$ as $z\rightarrow \infty$, but with an $R$-dependent power $\nu_R$. In the cases $R=1,2,3$ the characteristic polynomial of the transfer matrix is less than quintic, and we are able to extract this exponent
analytically, finding $\nu_1=1$, $\nu_2=2/3$, and $\nu_3=1/2$. Numerical studies suggest that this power decreases monotonically with increasing $R$.

\textit{Summary and outlook.}-- We introduced and studied the exact ground states of a class of Hamiltonians with ``blockade interaction''. We showed, among other results, that entanglement is only present within the blockaded region while non-classical correlations extend significantly further. One might speculate that this could find practical implications for the use of chains of Rydberg atoms or polar molecules as physical platforms for quantum information processing, communication or metrology \cite{Bose03}.

\begin{acknowledgements}
\textit{Acknowledgements}-- E. L. would like to thank M. Cianciaruso, G. Adesso and M. Marcuzzi for the insightful discussions. The research leading to these results has received funding from the European Research Council under the European Union's Seventh Framework Programme (FP/2007-2013) /
ERC Grant Agreement n. 335266 (ESCQUMA). We also acknowledge financial support from EPSRC Grant no.\ EP/J009776/1.
\end{acknowledgements}

%

\end{document}